# Emergent charge order near the doping-induced Mott-insulating quantum phase transition in $Sr_3Ru_2O_7$

Justin Leshen[1†], Mariam Kavai[1†], Ioannis Giannakis[1†], Yoshio Kaneko[2], Yoshi Tokura[2], Shantanu Mukherjee[1, 3], Wei-Cheng Lee[1], and Pegor Aynajian[1]*

[1]Department of Physics, Applied Physics and Astronomy, Binghamton University, Binghamton, NY
[2]RIKEN Center for Emergent Matter Science, Wako 351-0198 Japan
[3]Indian Institute of Technology, Madras, India

[†]These authors contributed equally to this work
*Corresponding author: aynajian@binghamton.edu

**We investigate the double layered $Sr_3(Ru_{1-x}Mn_x)_2O_7$ and its doping-induced quantum phase transition (QPT) from a metal to an antiferromagnetic (AFM) Mott insulator. Using spectroscopic imaging with the scanning tunneling microscope (STM), we visualize the evolution of the electronic states in real- and momentum-space. We find a partial-gap in the tunneling density of states at the Fermi energy ($E_F$) that develops with doping to form a weak Mott insulating ($\Delta \sim 100meV$) state. Near the QPT, we discover a spatial electronic reorganization into a commensurate checkerboard charge order. These findings share some resemblance to the well-established universal charge order in the pseudogap phase of cuprates. Our experiments therefore demonstrate the ubiquity of the incipient charge order that emanates from doped Mott insulators.**

Search for novel electronically ordered states of matter emerging near quantum phase transitions (QPT) is an intriguing frontier of condensed matter physics [1–3]. In Ruthenates, the interplay between Coulomb correlations among the 4d electronic states and their spin-orbit interactions, lead to complex forms of electronic phenomena. Quantum critical behavior, electronic nematic, and spin-triplet superconductivity are a few examples that unfold when these systems are non-thermally tuned [4–8].

The bilayer ruthenate, $Sr_3Ru_2O_7$, has a complex quasi-two dimensional electronic structure due to the rotation of the bulk $RuO_6$ octahedra that leads to the reconstruction of the Fermi surface (Fig.1(a)). As a result, multiple electronic bands cross the Fermi level, revealed by angle resolved photoemission spectroscopy (ARPES) [9] and de Haas-van Alphen (dHvA) [10] studies. While naively, the extended 4d Ru-orbitals as compared to 3d orbitals are expected to make it a weakly correlated metal, $Sr_3Ru_2O_7$ is one of the most strongly renormalized heavy d-electron material systems, demonstrated by the heavy flat electronic bands in previous ARPES [9] and STM [11] experiments. Signatures of the inherent electronic correlations are further emphasized by the impact of minute perturbations on its electronic ground state. The most prominent is the magnetic field-tuned quantum critical behavior and emergent electronic nematic order [7,8,12] with spin density wave instability [13] seen in transport and neutron scattering.

Doping acts as another pertinent non-thermal tuning parameter of the electronic states of $Sr_3Ru_2O_7$. At a few percent of Mn replacing Ru, $Sr_3(Ru_{1-x}Mn_x)_2O_7$ undergoes a metal to Mott-insulating QPT, accompanied at lower temperatures by an E-type AFM order with a wave vector $Q_{AFM} = (\pi/2a, \pi/2a)$ (Fig.1(b)) [14–16] analogous to the AFM structure of FeTe [17]. Resonant x-ray scattering (REXS) indicates the intensity and correlation length (not exceeding 160nm) of the AFM order to *decrease* at lower temperatures (T<<$T_{AFM}$), perhaps providing indications of a competing order [16,18]. While the origin of the insulating state remains unclear, the Coulomb correlations between the localized 3d Mn orbitals and the relatively extended 4d Ru orbitals may play a crucial role in the metal to insulator transition. X-ray diffraction indicates the $RuO_6$ octahedral-rotation to decrease with doping and be completely suppressed near x = 0.2 [19,20]. On the other hand, x-ray absorption experiments [21] reveal an unexpected $3^+$ valence of the Mn suggesting that Mn doping does not introduce holes to the Fermi surface, rather tunes the electronic states likely by structural distortions as well as enhanced Coulomb correlations with increased doping. Such a structural distortion clearly impacts the electronic structure and makes it susceptible towards magnetic instabilities [22]. Indeed, previous electron microscopy and spectroscopy experiments have shown, in real space, the extreme sensitivity of the Mott-type metal-insulator transition to applied mechanical stress/strain [23].

At the border of AFM and Mott insulating QPT, emergent electronic instabilities are frequently discovered. We use spectroscopic imaging with the STM to investigate the doping-induced metallic to AFM and Mott-insulating QPT in $Sr_3(Ru_{1-x}Mn_x)_2O_7$. A major goal is to explore whether broken electronic symmetry states emerge near the QPT. Figure 1(c) shows a topographic STM image of single crystal $Sr_3(Ru_{1-x}Mn_x)_2O_7$ with x=5% cleaved in-situ in our ultrahigh-vacuum STM exposing the SrO surface. The Fourier transform (FT) of the topograph reveals atomic Bragg peaks at the corners of the map corresponding to ($\pm 2\pi/a$, 0) and (0, $\pm 2\pi/a$) (Fig.1(d)). Note that, throughout this paper, we use tetragonal notation with lattice constant $a \sim 3.9$Å. Additional satellite reflections at ($\pm\pi/a, \pm\pi/a$) correspond to the doubling of the bulk unit cell owing to the rotational distortion of the $RuO_6$ octahedra (Fig.1(a),(d)) as previously observed [11,20]. The Mn dopants can be identified by their bias-dependent signatures in the topographic images (Fig.1(c)). They can be best visualized at a sample bias $V \sim 1$ V as bright spots for different doped samples (Fig.1(e)). Counting the number of Mn dopants on the surface reveals a systematic down shift (~ 2%) with a lower surface doping as compared to the expected bulk (see Supplementary Information).

The local impact of Mn-dopants on the spatial electronic structure, over a range of doping spanning the QPT, can be visualized through the spectroscopic imaging shown in Fig.2. At low Mn concentration (x=1%; much below the QPT) the local electronic density of states (LDOS) reveal a particle-hole asymmetric V-shaped partial-gap of ~ ±30 meV near the Fermi energy at T = 13 K (Fig.2(a)-(c)). The density of states is similar to what has been previously observed in the undoped parent compound [24]. In particular, a small hump at $E_F$ previously identified as a van-Hove singularity [9, 24], is also present in our spectra obtained away from Mn dopants (Fig.2(c) and supplementary information). The effect of the Mn, at this low concentration, shows modifications of the local gap predominantly near $E_F$ (see Fig.2(b),(c) and see supplementary information). Conductance maps $dI/dV$ (**r**,$V$) $\equiv$ g(**r**, $V$) locally at locations **r** and at a constant sample bias $V$, displayed in Fig.2(b), reveal that, in this dilute limit the Mn-dopants spatially modify the LDOS within a length scale of 20Å.

Increasing the Mn-doping towards the QPT reveals a striking spatial reorganization of the local electronic states. Figure 2(d)-(f) display a topographic image and the LDOS, showing the evolution of the partial-gap in the x=5% sample. Conductance maps g(**r**, $V$) reveal a bi-directional long-range checkerboard charge modulations (Fig.2(e)); particularly V=-40meV and V=-100meV maps) embedded within the long-wavelength inhomogeneity. The absence of this charge ordering in the real space images of the x=1% sample (Fig.2(b)), as well as in earlier STM experiments on the undoped [11,24] and Mn-doped compound at 100 K [20], indicate its origin not to be related to any surface reconstruction, rather to an electronic instability induced by Mn-doping at low temperatures. Further indication of the electronic nature of the ordering phenomenon can be concluded by the absence of any structural peaks, corresponding to the observed modulations, in neutron scattering [14], in recent low energy electron diffraction (LEED) [20], and in non-resonant hard x-ray scattering [18]. The observed charge modulations, along the Ru-O bond direction with a periodicity of 2$a$ (corresponding to a wave vector q =

($\pm\pi/a$, 0) and (0, $\pm\pi/a$) ≡ Q*) therefore corroborate an emergent electronic charge ordering near the metal-to-Mott-insulator phase transition.

At even higher doping (x=10%), the STM spectra reveals strong inhomogeneity with areas of V-shaped partial gaps mixed with a complete, yet inhomogeneous, gapping of the Fermi surface into a U-shaped insulating gap of ~100meV (Fig.2(i) and supplementary). This is indicative of a close proximity to the metal-to-Mott insulator transition [25]. The evolution of the gap with doping resembles the opposite trend of what is seen in the hole-doped, spin-orbit driven, Mott insulator $Sr_3(Ir_{1-x}Ru_x)_2O_7$ [25]. The conductance maps g(**r**, *V*) (fig.2(h)) reveal inhomogeneous short-range modulations, equally populated along the two atomic directions with their periodicity locked to 2*a*. The glassy nature of the observed charge order could be a consequence of the larger electronic inhomogeneity introduced by the Mn-dopants and the transition into the Mott insulating state.

It is important to emphasize that the doping-induced metal to insulator transition evaluated from the surface LDOS at T = 13 K (Blue points in Fig.1(b)) closely follows that extracted from bulk resistivity measurements (gray area in Fig.1(b)) [16]. The slight offset may be a result of the lower doping on the surface as compared to the bulk. This may also explain the slight enhancement of the octahedral distortion on the surface of $Sr_3(Ru_{1-x}Mn_x)_2O_7$ as compared to the bulk [26].

To visualize the momentum-space structure of these modulations and their doping dependence, we display in Fig.3 the real-space conductance maps and selected FTs at several energies taken on a larger surface area for optimal momentum-space resolution. For x=1%, in addition to the Bragg peaks at ($\pm 2\pi/a$, 0) and (0, $\pm 2\pi/a$) and the octahedral rotation satellite peaks at ($\pm\pi/a$, $\pm\pi/a$), the FTs show dispersive features ($q_2$, $q_7$ and $q_8$) along the high symmetry directions, corresponding to quasiparticle interference (Fig.3(a), (b)) (here we follow the same notations as reference [11], similar to those previously seen in the STM studies of 1% Ti-doped $Sr_3Ru_2O_7$ [11]. The extracted dispersions along ($\pm\pi/a$, $\pm\pi/a$) are shown in Fig.3(g). While no clear peak is observed at the Q* location in this 1% sample, there is a faint and broad intensity in the FT around this wave vector ($q_1$ in Fig.3b, whose dispersion is difficult to extract due to its proximity to $q_7$ and their weak intensity), which locally originates only from areas with Mn dopants (more detail in supplementary information).

In contrast, for x=5%, the FTs indicate distinct sharp and non-dispersive peaks at Q* = ($\pm\pi/a$, 0) and (0, $\pm\pi/a$) that correspond to a real-space modulation with a periodicity of 2*a* (Fig.3(d),(h)). In addition to the Q* modulations, we also observe strong non-dispersive modulations coinciding with the AFM ordering wavevector, $Q_{AFM}$ = ($\pm\pi/2a$, $\pm\pi/2a$) (Fig.3(d),(g)). In principle, without a spin-polarized tip, STM is not sensitive to a magnetic contrast. Our experiments indicate that the AFM order also has a charge component, which was suggested by previous REXS experiments [16]. Moving into the Mott state

(x=10%) reveals the Q* modulations to broaden, yet dominate and remain locked to the same commensurate momentum, independent of doping (Fig.3(f),(h)).

These experiments, unsurprisingly, indicate strong interplay between spin and charge instabilities. Close to the QPT (x ~ 5%) the system is susceptible to long-range charge ordering at Q* and $Q_{AFM}$ (Fig.3) as well as a competing short-range AFM ordering at $Q_{AFM}$ (as revealed by previous REXS [16,18]). Signs of such a competition can also be seen in the suppressed AFM correlation length with decreasing temperature in REXS [16,18]. This presents an exotic electronic phase in which charge and spin orders simultaneously coincide at the same wavevector. With further doping, however, the charge ordering dominates at Q*, whereas a long-range E-type AFM order (with negligible charge contribution, as seen in our STM data (Fig.3(f)) dominates at $Q_{AFM}$ [14,18]. It is important to note that although STM is primarily sensitive to charge order, some forms of orbital orders can also be detected through the STM tip sensitivity to different orbital wave functions [27,28] and therefore orbital contribution to $Q_{AFM}$ cannot be omitted as a possible scenario.

To provide insights into the origins of the emergent broken symmetry states at Q* and $Q_{AFM}$, we compute the bare susceptibility $\chi(q,\omega)$ using a 12 orbital tight-binding model [29], fit to match the Fermi surface measured by ARPES on $Sr_3Ru_2O_7$ (see methods and supplementary information). The calculations of the bare susceptibility, $\chi(q,0)$, whose peaks correspond to the wavevectors with strong Fermi surface nesting, reveal broad features around Q* = ($\pi/a$, 0) and $Q_{AFM}$ = ($\pi/2a$, $\pi/2a$), which are enhanced with increased Mn-doping (see supplementary information). As previously shown, the effect of Mn-dopants on the $Sr_3Ru_2O_7$ band structure can be partly accounted for by the suppression of the $RuO_6$ octahedral rotation in the tight-binding band structure [31]. As doping increases, the suppressed rotational-distortion (modeled as suppression in the hopping parameter [31]) reduces the coupling between the $d_{xz}$ and $d_{yz}$ orbitals making the $\alpha_2$ band more square-like. This itself enhances the nesting at Q* and $Q_{AFM}$, leading to an increase in $\chi(q,0)$ (see supplementary information). Our theoretical analysis therefore indicates that a *broad* Fermi surface nesting does exist near Q* and $Q_{AFM}$ near x ~ 5%. However, we would like to emphasize that nesting alone cannot explain the emergent charge order at Q* and it will require enhanced anisotropic electron-phonon (*e-p*) coupling and/or strong coulomb correlations to induce a commensurate charge ordering instability. Recently, strong *e-p* coupling have been observed at the charge ordering wavevectors in $YBa_2Cu_3O_{6.6}$ [32], $Bi_2Sr_2CaCu_2O_{8+x}$ [33] and $NbSe_2$ [34]. In addition, a significant softening in the temperature dependence of the $B_{1g}$ phonon mode has also been observed to coincide with the metal to Mott insulator transition in $Ca_3Ru_2O_7$ pointing towards the possibility of similar physics in Mn-doped $Sr_3Ru_2O_7$. Inelastic resonant x-ray or neutron scattering [35,36] can probe the momentum dependence of the *e-p* coupling and such future experiments may offer a more complete picture of the origin of the observed instabilities in $Sr_3(Ru_{1-x}Mn_x)_2O_7$.

Finally, we examine the structure and symmetry of the charge ordering at Q*. Figure 4(a) shows a zoom-in of the conductance map for the x=10% sample (white box in Fig.3(e)). While the charge order can be clearly observed, the long-wavelength inhomogeneity in the conductance map obscures the detail of the ordering patterns. To visualize the real space character of these patterns, we Gaussian-filter the Q* peaks in the raw unsymmetrized FTs (see supplementary information) and inverse Fourier transform back into real space. Figure 4(b) shows the charge order map corresponding to Fig.4(a). The figure displays a mixture of spatially inhomogeneous checkerboard and stripe patterns. To quantitatively distinguish between the two glassy orders, we compute the amplitudes of the two orthogonal components of the charge order (Fig.4(c), (d)) and calculate their cross-correlation (Fig.4(e)) (see Supplementary Information). In a checkerboard order, the two orthogonal components are locally correlated (positive cross-correlation). A stripe order leads to a local anti-correlation between the two orthogonal charge ordering amplitudes. The positive cross-correlation observed as a function of bias (Fig.4(e)) indicates globally a four-fold symmetric bi-directional charge order.

In conclusion, our experiments reveal non-dispersive commensurate charge order instability to emerge near the metal to AFM Mott insulator transition in $Sr_3Ru_2O_7$. These findings draw similarities to the ubiquitous charge order seen in STM and x-ray experiments on cuprates [37–44]. While the metal to Mott insulator transition in $Sr_3Ru_2O_7$, which is likely related to structural distortion and enhanced Coulomb correlations by 3d Mn replacing 4d Ru, is quite different than the charge doping-induced Mott insulator to metal transition in cuprates, it is quite intriguing that a charge ordered state with similar real-space patterns exists near the Mott transition of both systems. Finally, similar charge modulation patterns on strong spin-orbit coupled $Sr_2IrO_4$ have recently been found near its doping-induced Mott insulator to metal transition with a similar periodicity to our Q* [45]. Overall, all these experiments, independent of their detailed intrinsic electronic structure, point to ubiquitous tendencies towards some form of charge order emerging at the border of Mott insulators.

**Methods:**

**Sample preparation and characterization:**

The single crystals of $Sr_3(Ru_{1-x}Mn_x)_2O_7$ used in our experiments were grown at RIKEN laboratory using the floating zone technique. These single crystals were obtained under 1M Pa. atmosphere of argon gas containing 10% oxygen. Small flat samples were cut down to ~ (2 x 2 x 0.5mm$^3$) and attached to aluminum plates using H20E conducting epoxy. Aluminum cleaving posts were then attached to samples perpendicular to the a-b cleaving plane using H74F non conducting epoxy. Samples were cleaved at room temperature in ultrahigh vacuum by knocking the post off, and then immediately transferred in situ to the custom RHK Pan STM head, which had been cooled down to T ~ 13 K. STM topographs were taken in constant current mode and dI/dV measurements were performed using a standard lock-in technique with a reference frequency of 0.921 kHz. PtIr tips were used in all experiments. Tips were prepared prior to each experiment on a Cu(111) surface that had been treated with several rounds of sputtering and annealing, and then placed into the microscope head to cool. The sample and Cu were placed next to each other inside the microscope head so as to reduce exposure and preserve tip structure when moving from one to the other. The data presented in this paper were collected from approximately 3 successfully cleaved samples for each doping percentage (1%, 5%, 7.5%, and 10%). There were negligible differences measured between different cleaves and between different areas on a sample.

**Tight binding calculations:**

To calculate the bare susceptibility, we start from a 12 orbital tight-binding model that contains $t_{2g}$ orbitals, the bilayer structure, and the spin-orbit coupling. The hopping parameters are determined by fitting the Fermi surfaces observed in ARPES. Details can be found in Ref. [29]. The resulting Hamiltonian in the momentum space can be expressed as

$$H_0 = \sum_{s=\uparrow,\downarrow} \sum_{k_z=0,\pi} \sum_k \Phi_k^+ h_k(k_z) \Phi_k$$

where $\Phi_k = \left(d_{k,s,k_z}^{yz}, d_{k,s,k_z}^{xz}, d_{k,-s,k_z}^{xy}, d_{k+Q,s,k_z}^{yz}, d_{k+Q,s,k_z}^{xz}, d_{k+Q,-s,k_z}^{xy}\right)$ are the electron annihilation operators, $k$ is the in-plane momentum, $k_z$ is the out-of-plane momentum associated with the bilayer structure, $Q=(\pi, \pi)$ is the wavevector associated with the octahedral rotation, and $s$ is the spin index. The matrix element $h_k(k_z)$ can be found in Ref. [29]. The retarded bare susceptibility in a multiorbital system can in general be written as a tensor of

$$\chi_{l_1,l_2,l_3,l_4}^{s,s'}(q,\omega) = \frac{1}{N} \sum_{k\alpha\beta} W_{l_1,l_2,l_3,l_4}^{\alpha,\beta} \frac{n_F\left(E_\beta(k+q)\right) - n_F\left(E_\alpha(k)\right)}{\omega + i\delta + E_\alpha(k) - E_\beta(k+q)}$$

where $W_{l_1,l_2,l_3,l_4}^{\alpha,\beta} = \psi_{l_1s}^{\alpha*}(k)\psi_{l_2s'}^{\beta}(k+q)\psi_{l_3s'}^{\beta*}(k+q)\psi_{l_4s}^{\alpha}(k)$, and $\psi_{ls}^{\alpha}(k)$ is the component of the eigenvector of the eigenstate $\alpha$ projected on the state with orbital $l$ and spin $s$. This bare susceptibility expressed in the form of Lindhard function with extra matrix elements takes into account the scatterings between different orbitals in a 12 orbital model. To perform the momentum sum over the Lindhard function, we have used a k-mesh of (Nx × Ny=250 × 250 points) in all the calculations. In the normal state, the system still has the spin rotational symmetry in the presence of the spin orbit coupling. As a result, $\chi_{l_1,l_2,l_3,l_4}^{s,s'}(q,\omega)$ is the same for any combination of (s, s'), and consequently we can omit (s, s'). Finally, we compute

$$\chi(q,0) = \sum_{l_1,l_2,l_3,l_4} \chi_{l_1,l_2,l_3,l_4}(q,0)$$

which exhibits peaks at wavevectors corresponding to the Fermi surface nesting.

As the doping is introduced, the octahedral rotation is reduced. The leading effect of this octahedral rotation is to enable hopping between different orbitals on nearest neighbor sites. The details of this model is discussed in Ref. [29]. The reduced octahedral rotation results in less hybridization between the $d_{xz}$ and $d_{yz}$ orbitals. Note that the current model cannot capture the insulating behavior in the paramagnetic state.

**References:**

[1] Vojta, M. Quantum phase transitions. Reports Prog. Phys. **66**, 2069 (2003).

[2] Sachdev, S. & Keimer, B. Quantum criticality. Phys. Today **64**, 29 (2011).

[3] Dagotto, E. Complexity in strongly correlated electronic systems. Science **309**, 257 (2005).

[4] Behrmann, M., Piefke, C. & Lechermann, F. Multiorbital physics in Fermi liquids prone to magnetic order. Phys. Rev. B **86**, 045130 (2012).

[5] Mackenzie, A. P. & Maeno, Y. The superconductivity of $Sr_2RuO_4$ and the physics of spin-triplet pairing. Rev. Mod. Phys. **75**, 657 (2003).

[6] Berridge, A. M., Green, A. G., Grigera, S. A. & Simons, B. D. Inhomogeneous magnetic phases: A Fulde-Ferrell-Larkin-Ovchinnikov-like phase in $Sr_3Ru_2O_7$. Phys. Rev. Lett. **102**, 136404 (2009).

[7] Rost, A. W., Perry, R. S., Mercure, J. F., Mackenzie, A. P. & Grigera, S. A. Entropy Landscape of Phase Formation Associated with Quantum Criticality in $Sr_3Ru_2O_7$. Science **325**, 1360 (2009).

[8] Grigera, S. A. et al. Magnetic field-tuned quantum criticality in the metallic ruthenate $Sr_3Ru_2O_7$. Science **294**, 329 (2001).

[9] Tamai, A. et al. Fermi surface and van hove singularities in the itinerant metamagnet $Sr_3Ru_2O_7$. Phys. Rev. Lett. **101**, 026407 (2008).

[10] Borzi, R. A. et al. de Haas-van Alphen effect across the metamagnetic transition in $Sr_3Ru_2O_7$. Phys. Rev. Lett. **92**, 216403 (2004).

[11] Lee, J. et al. Heavy d-electron quasiparticle interference and real-space electronic structure of $Sr_3Ru_2O_7$. Nat Phys **5**, 800 (2009).

[12] Rost, A. W. et al. Thermodynamics of phase formation in the quantum critical metal $Sr_3Ru_2O_7$. Proc. Natl. Acad. Sci. USA **108**, 16549 (2011).

[13] Lester, C. et al. Field-tunable spin-density-wave phases in $Sr_3Ru_2O_7$. Nat. Mater. **14**, 373 (2015).

[14] Mesa, D. et al. Single-bilayer E-type antiferromagnetism in Mn-substituted $Sr_3Ru_2O_7$: Neutron

[15] Mathieu, R. et al. Impurity-induced transition to a Mott insulator in $Sr_3Ru_2O_7$. Phys. Rev. B **72**, 92404 (2005).

[16] Hossain, M. A. et al. Mott versus Slater-type metal-insulator transition in Mn-substituted $Sr_3Ru_2O_7$. Phys. Rev. B **86**, 41102 (2012).

[17] Bao, W. et al. Tunable (δπ, δπ)-Type Antiferromagnetic Order in α-Fe(Te,Se) Superconductors. Phys. Rev. Lett. **102**, 247001 (2009).

[18] Hossain, M. A. et al. Electronic superlattice revealed by resonant scattering from random impurities in $Sr_3Ru_2O_7$. Sci. Rep. **3**, 2299 (2013).

[19] Hu, B. et al. Structure-property coupling in $Sr_3(Ru_{1-x}Mn_x)_2O_7$. Phys. Rev. B **84**, 174411 (2011).

[20] Li, G. R. et al. Atomic-Scale Fingerprint of Mn Dopant at the Surface of $Sr_3(Ru_{1-x}Mn_x)_2O_7$. Sci. Rep. **3**, 2882 (2013).

[21] Hossain, M. A. et al. Crystal-field level inversion in lightly Mn-doped $Sr_3Ru_2O_7$. Phys. Rev. Lett. **101**, 016404 (2008).

[22] Fang, Z. & Terakura, K. Magnetic phase diagram of $Ca_{2-x}Sr_xRuO_4$ governed by structural distortions. Phys. Rev. B **64**, 20509 (2001).

[23] Kim, Tae-Hwan. et al. Imaging and manipulation of the competing electronic phases near the Mott metal-insulator transition. PNAS. 107, 5272 (2010).

[24] Iwaya, K. et al. Local tunneling spectroscopy across a metamagnetic critical point in the bilayer ruthenate $Sr_3Ru_2O_7$. Phys. Rev. Lett. **99**, 057208 (2007).

[25] Dhital, C. et al. Carrier localization and electronic phase separation in a doped spin-orbit-driven Mott phase in $Sr_3(Ir_{1-x}Ru_x)_2O_7$. Nat. Commun. **5**, 3377(2014).

[26] Hu, B. et al. Surface and bulk structural properties of single-crystalline $Sr_3Ru_2O_7$. Phys. Rev. B **81**, 184104 (2010).

[27] Aynajian, P. et al. Visualizing Heavy Fermion Formation and their Unconventional Superconductivity in f-Electron Materials. J. Phys. Soc. Japan **83**, 61008 (2014).

[28] Aynajian, P. et al. Visualizing heavy fermions emerging in a quantum critical Kondo lattice. Nature **486**, 201 (2012).

[29] Lee, W. C., Arovas, D. P. & Wu, C. Quasiparticle interference in the unconventional metamagnetic compound $Sr_3Ru_2O_7$. Phys. Rev. B **81**, 184403 (2010).

[30] Capogna, L. et al. Observation of two-dimensional spin fluctuations in the bilayer ruthenate $Sr_3Ru_2O_7$ by inelastic neutron scattering. Phys. Rev. B **67**, 012504 (2003).

[31] Mukherjee, S. & Lee, W.C. Structural and magnetic field effects on spin fluctuations in $Sr_3Ru_2O_7$. Phys. Rev. B **94**, 64407 (2016).

[32] Tacon, M. Le. et al. Inelastic X-ray scattering in YBa$_2$Cu$_3$O$_{6.6}$ reveals giant phonon anomalies and elastic central peak due to charge-density-wave formation. Nat. Phys. **10**, 52 (2013).

[33] Chaix, L. et al. Dispersive charge density wave excitations in Bi$_2$Sr$_2$CaCu$_2$O$_{8+\delta}$. Nat. Phys. **13,** 952 (2017).

[34] Weber, F. et al. Extended phonon collapse and the origin of the charge-density wave in 2H-NbSe$_2$. Phys. Rev. Lett. **107**, 107403 (2011).

[35] Aynajian, P. et al. Energy gaps and Kohn anomalies in elemental superconductors. Science **319**, 1509 (2008).

[36] Keller, T. et al. Momentum-resolved electron-phonon interaction in lead determined by neutron resonance spin-echo spectroscopy. Phys. Rev. Lett. **96**, 225501 (2006).

[37] Hoffman, J. E. et al. A Four Unit Cell Periodic Pattern of Quasi-Particle States Surrounding Vortex Cores in Bi$_2$Sr$_2$CaCu$_2$O$_{8+\delta}$. Science **295**, 466 (2002).

[38] Parker, C. V. et al. Fluctuating stripes at the onset of the pseudogap in the high-Tc superconductor Bi$_2$Sr$_2$CaCu$_2$O$_{8+x}$. Nature **468**, 677 (2010).

[39] Ghiringhelli, G. et al. Long-Range Incommensurate Charge Fluctuations in (Y,Nd)Ba$_2$Cu$_3$O$_{6+x}$. Science **337**, 821 (2012).

[40] Chang, J. et al. Direct observation of competition between superconductivity and charge density wave order in YBa$_2$Cu$_3$O$_{6.67}$. Nat. Phys. **8**, 871 (2012).

[41] Tabis, W. et al. Charge order and its connection with Fermi-liquid charge transport in a pristine high Tc cuprate. Nat. Commun. **5**, 5875 (2014).

[42] da Silva Neto, E. H. et al. Ubiquitous Interplay Between Charge Ordering and High-Temperature Superconductivity in Cuprates. Science **343**, 393 (2014).

[43] Comin, R. & Damascelli, A. Resonant X-Ray Scattering Studies of Charge Order in Cuprates. Annu. Rev. Condens. Matter Phys. **7**, 369 (2016).

[44] da Silva Neto, E. H. et al. Charge ordering in the electron-doped superconductor Nd$_{2-x}$Ce$_x$CuO$_4$. Science **347**, 283 (2015).

[45] Battisti, I. et al. Universality of pseudogap and emergent order in lightly doped Mott insulators. Nat. Phys. **13**, 21 (2016).

[46] Momma, K. & Izumi, F. VESTA 3 for three-dimensional visualization of crystal, volumetric and morphology data. J. Appl. Crystallogr. **44**, 1272 (2011).

**Acknowledgments**

We thank Michael Lawler for helpful discussions. P.A. thanks Abhay Pasupathy for helpful technical discussions.

P.A. acknowledges support from the U.S. National Science Foundation (NSF) CAREER under award No. DMR-1654482. S.M. acknowledges support from New Faculty Initiative Grant from IIT Madras under Project No. PHY/16-17/865/NFIG/SHAA.

**Author Contributions:**

J.L., M.K., and I.G. performed the STM measurements and data analysis. Y.K. and Y.T. synthesized and characterized the materials. S.M. and W.C.L carried out the theoretical modelling and analysis. P.A. wrote the manuscript. All authors commented on the manuscript.

**Fig.1**

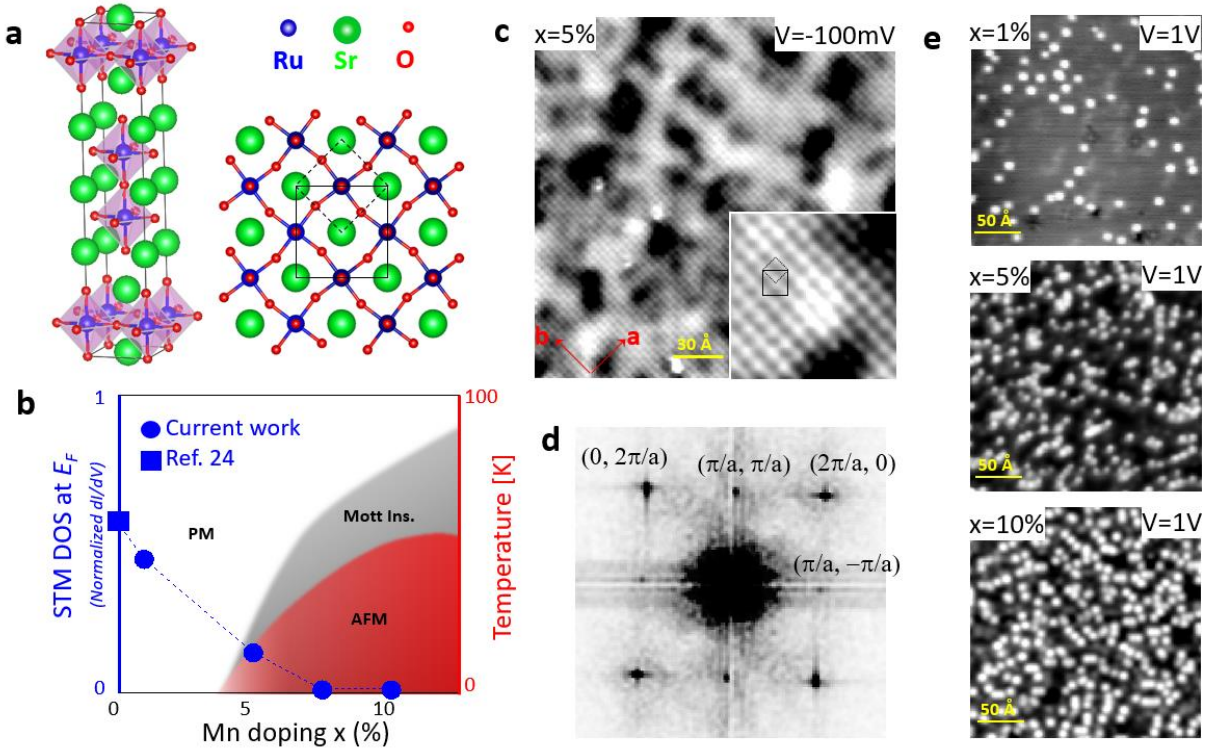

**FIG. 1**: **Structure and phase diagram of $Sr_3(Ru_{1-x}Mn_x)_2O_7$.** (**a**), Left: Crystal structure of $Sr_3(Ru_{1-x}Mn_x)_2O_7$. Right: Top view of the SrO cleaving plane along with the sublayer $RuO_6$ octahedra [46]. The Mn dopants replace the Ru atoms at the center of the octahedra. The dashed square shows the (1 x 1) tetragonal unit cell, while the larger solid square shows the (√2 x √2)R45° unit cell created by the rotational distortion of the octahedra. (**b**), Schematic phase diagram of $Sr_3(Ru_{1-x}Mn_x)_2O_7$ showing the doping-temperature (right-axis) transition from a PM metal to an AFM Mott insulator [16]. The blue circles represent the doping dependence of the STM LDOS at $E_F$ measured at T=13 K (left axis) normalized to value at 200meV. The blue square is extracted from ref. [24]. (**c**), Constant current topographic image of a 5% Mn sample taken at -100mV and -500pA showing an atomically ordered surface. The inset shows a magnified image of the same area. (**d**), FT (plotted as the modulus) of the topograph in (**c**). Both the Bragg peaks at ($\pm 2\pi/a$, 0) and (0, $\pm 2\pi/a$) and satellite reflections at ($\pm\pi/a$, $\pm\pi/a$) are seen. (**e**), Topographic image taken on the x=1% (1V, 150pA), x=5% (1V, 150pA), and x=10% (1V, 2500pA) samples respectively. At higher bias voltages the Mn impurity sites appear as bright spots showing the progression of impurity density from 1% to 10%.

**Fig.2**

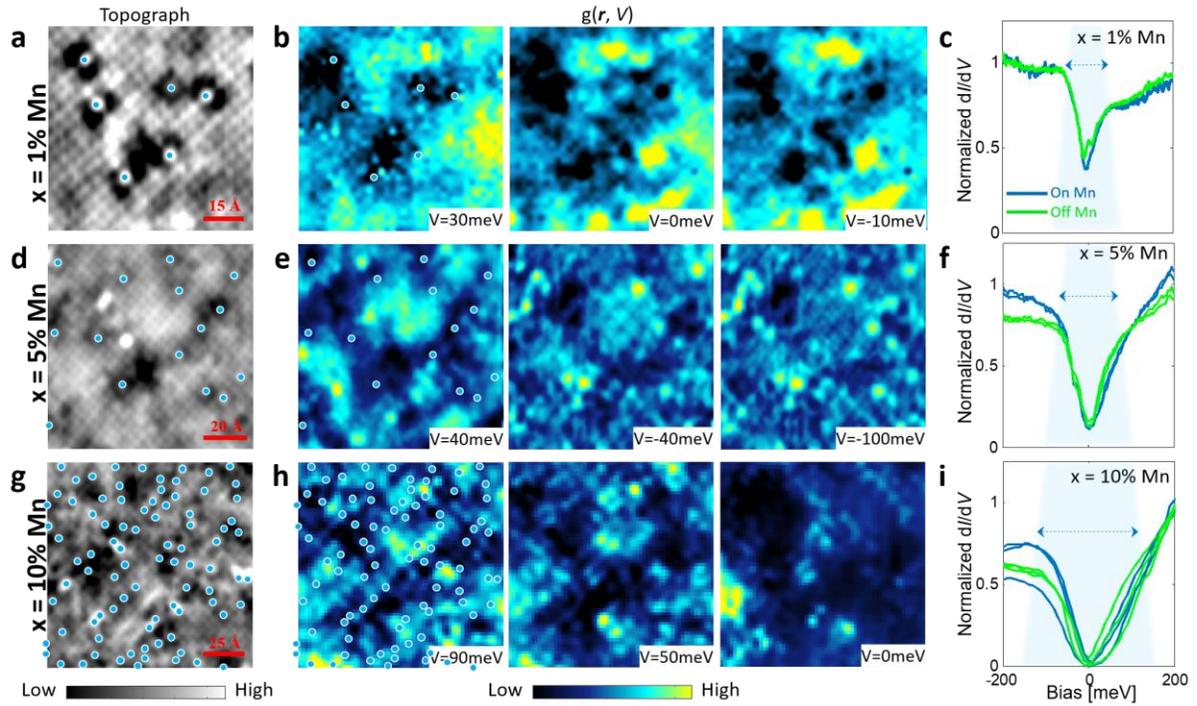

**FIG. 2**: **Real space structure of electronic states in $Sr_3(Ru_{1-x}Mn_x)_2O_7$.** Topographic image for (**a**), x=1% (150mV, 500pA), (**d**), x = 5% (200mV, 500pA), and (**g**), x = 10% (300mV, 2500pA) samples. Mn impurity sites are marked with a blue circle. Conductance maps at selected energies for (**b**), x=1%, (**e**), x = 5%, and (**h**), x = 10% samples taken on the same area as their corresponding topographs (with the same bias voltage and setpoint current). STM spectra taken on the on and off of Mn impurity sites for (**c**), x=1%, (**f**), x = 5%, and (**i**), x = 10% samples. The blue arrows emphasize the widening of the gap around $E_F$ with increased doping. All measurements were taken at T = 13 K.

Fig.3

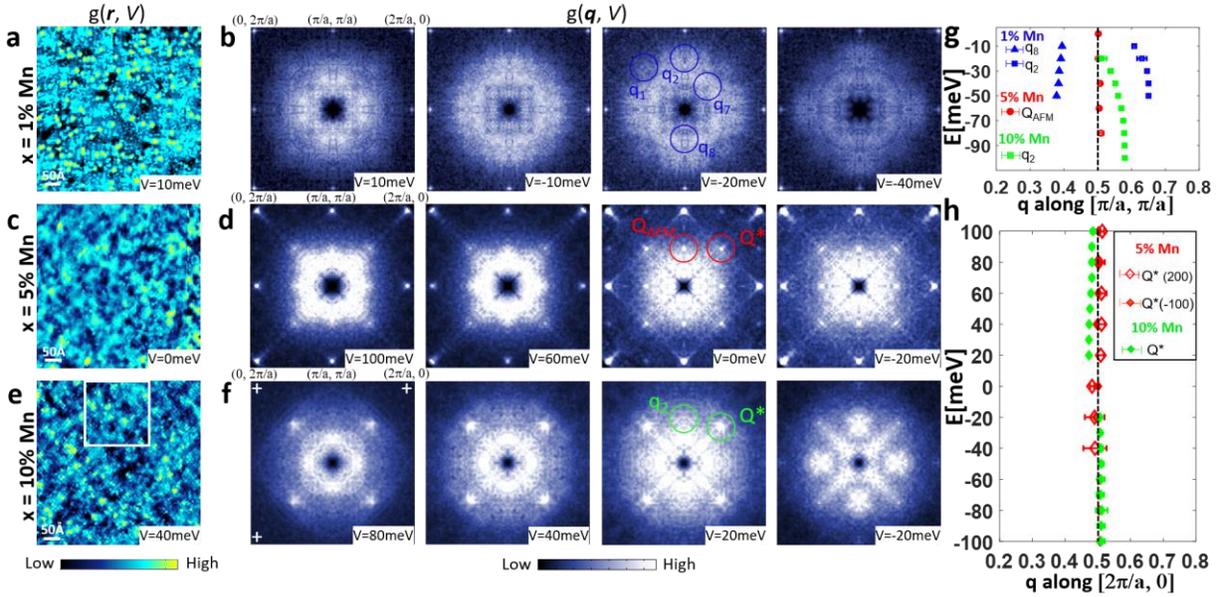

**FIG. 3: Momentum space structure of electronic states in $Sr_3(Ru_{1-x}Mn_x)_2O_7$**. Real space conductance maps and corresponding FTs (plotted as the modulus) at selected energies measured on **(a), (b)**, x = 1% (150mV, 500pA), **(c), (d)**, x = 5% (-100mV, -500pA), and **(e), (f)**, x = 10% (100mV, 2500pA) Mn doped samples. The Q* = (±π/a, 0) and (0, ±π/a) peak, $Q_{AFM}$ = (π/a, π/a), and QPI peaks $q_1$, $q_2$, $q_7$, and $q_8$ are all circled in the FTs. For the 10% sample, the Bragg peak intensities are too weak at these energies (inside the gap) to be seen but their location is marked by a white cross. All FTs are four-fold symmetrized but all the structures are visible in the unsymmetrized data (See supplementary information). All measurements were taken at T = 13 K. **(g),(h)** Energy dependency of the *q* value of the peaks and features for the 1%, 5%, and 10% samples. For the 5%Mn sample, Q* was extracted from two maps with a set-point bias of -100meV ( solid red diamonds) and 200meV (open red diamonds). The *q* values and error bars are extracted from the fits (see Supplementary Information for details).

**Fig.4**

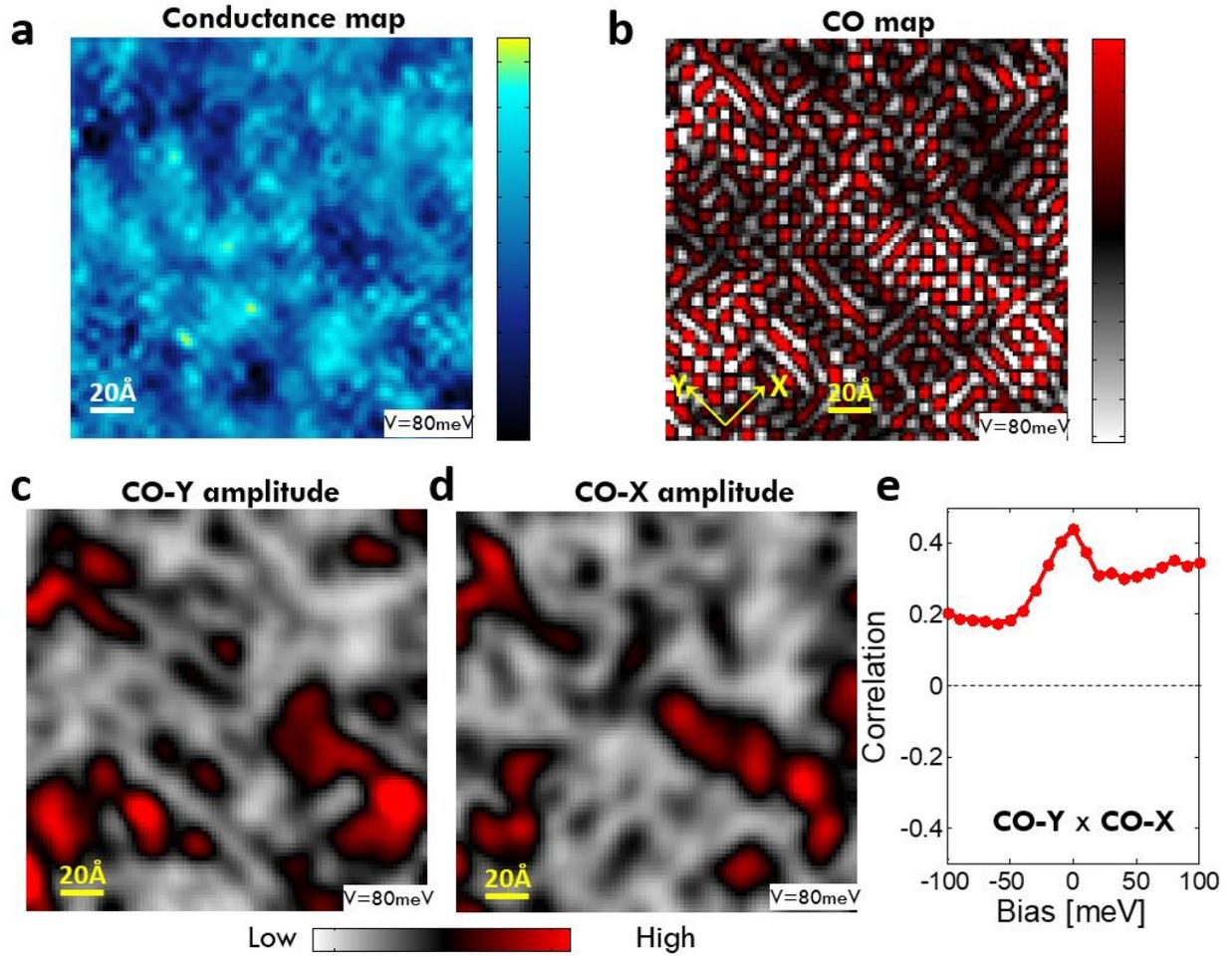

**FIG.4: Real-space structure of the emergent charge order**. (**a**), Real space conductance map for x=10% sample, corresponding to the area in the white box of Fig.3(e). (**b**), Inverse Fourier transform of the FT (preserving the phase) for x=10% sample after isolating the Q* peaks using a Gaussian profile filter (see supplementary information). The process enhances the glassy checkerboard charge order seen in (**a**) by suppressing the long wavelength inhomogeneity. (**c**), (**d**), Spatial amplitude of the charge order along the two orthogonal direction (Y and X), respectively. Higher intensity correspond to higher strength of the charge order along the specific (Y or X) direction. (**e**), Cross-correlation of the two orthogonal charge order amplitude maps (see supplementary information) as a function of bias. The positive correlation is an indication of a glassy checkerboard order.

# Supplementary Information for "Emergent charge order near the doping-induced Mott-insulating quantum phase transition in $Sr_3Ru_2O_7$"

Justin Leshen[1], Mariam Kavai[1], Ioannis Giannakis[1], Yoshio Kaneko[2], Yoshi Tokura[2], Shantanu Mukherjee[1,3], Wei-Cheng Lee[1], and Pegor Aynajian[1]*

[1]Department of Physics, Applied Physics and Astronomy, Binghamton University, Binghamton, NY

[2]RIKEN Center for Emergent Matter Science, Wako 351-0198 Japan

[3]Indian Institute of Technology, Madras, India

*Corresponding author: aynajian@binghamton.edu

## 1. Identifying the Mn dopants and density:

Identifying exact Mn dopant sites is important to many aspects of our analysis. The sample cleaves along the Sr-O plane revealing a layer with the Sr atoms appearing as a bright lattice of atoms and the apical Oxygen and sublayer Ru appearing as the lower intensity areas between the Sr lattice. The Mn atoms replace the Ru, which means they are expected to appear in between the surface atoms (Fig. S1). Figure S1 shows the impurity sites appearing as a bright spot centered over the Ru/Mn depression that is in between the surface Sr atoms as expected.

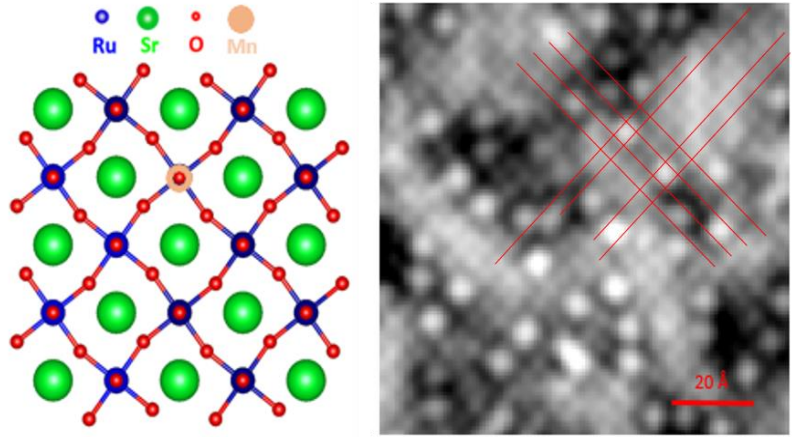

FIG. S1: **Lattice structure of $Sr_3(Ru_{1-x}Mn_x)_2O_7$** with one Mn atom shown replacing the Ru atom (left). STM topography of 10% Mn doped sample taken at $V_{bias}$= -0.3V, Setpoint=-0.5nA, and T=13K (right). The Mn dopant sites appear as bright spots centered above the Ru/Mn site. The red lines are a guide for the eye along the direction of the tetragonal lattice. The bright spots are centered between the four Sr atoms, which make up the tetragonal unit cell, exactly where the Mn atom should be based on the lattice structure.

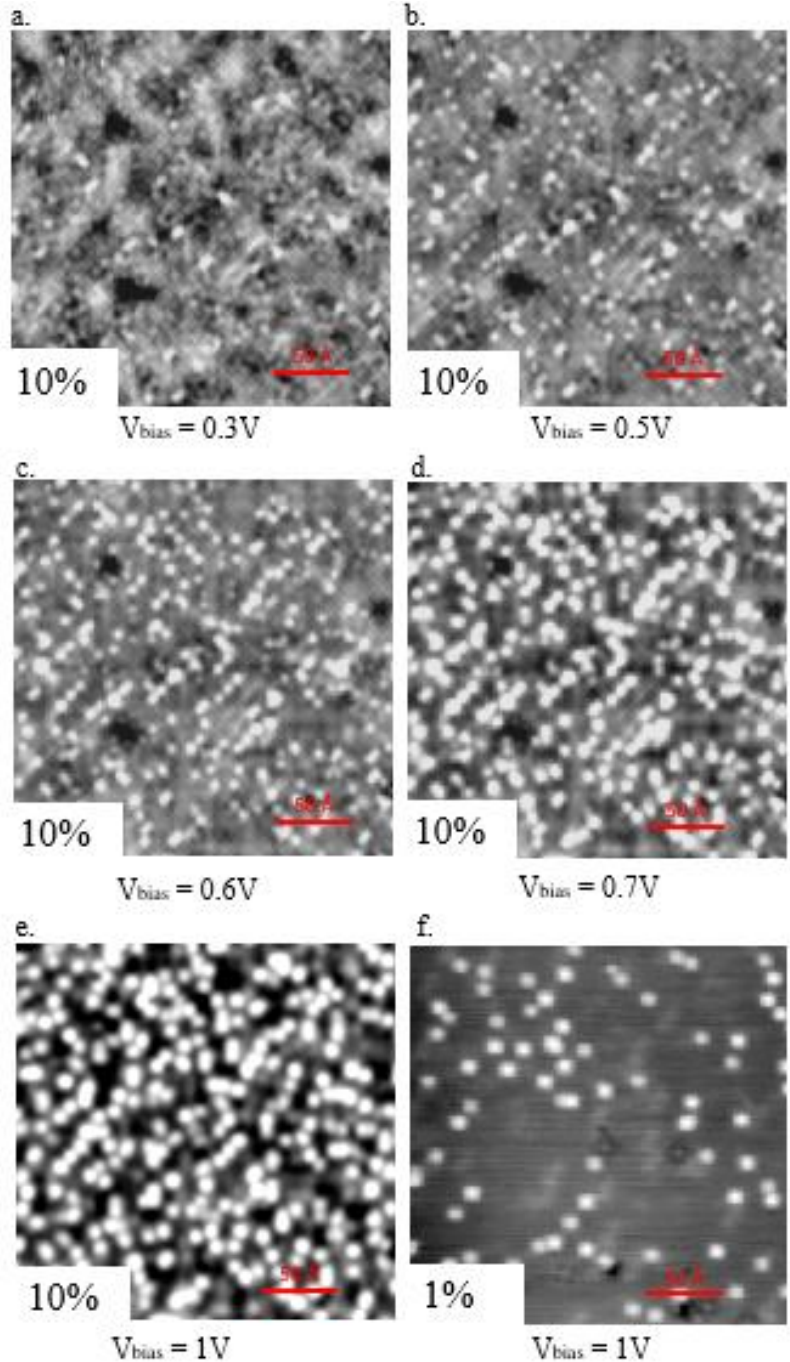

FIG. S2: **Evolution of dopants with setpoint bias.** (a)-(e). STM topographies taken on a 10% Mn sample, taken at the same field of view consecutively with only the bias voltage changed between them. The impurity sites change from bright spots surrounded by a dark depression, to just the bright spots between 0.3V and 0.5V, then gradually grow in size and intensity. The topographies were taken with a setpoint current of 2.5nA at T=13K and are 25nm x 25nm in size. (e)-(f) Show a 10% Mn and a 1% Mn sample respectively taken at a 1V bias on a same size area, showing the increase in the density of Mn in the sample.

There is a strong bias dependence on the appearance of the Mn dopants. Figure S2(a)-(e) shows several topographies taken on a 10% Mn doped sample, on the same field of view, with increasing bias. The Mn impurities appear as bright spots centered in between the surface atoms. As the bias is increased, the intensity of the bright spots increases until they dominate the image around 1V. Figure S2(f) shows a similar appearance for the Mn impurity at 1V in the 1% sample and a comparison between e and f shows how the impurity density increases between the 1% and 10% samples.

Counting these impurities on several topographies of varying sizes reveals a Mn percentage of $(1.3 \pm 0.1)$% for the 1%, $(2.6 \pm 0.5)$% for the 5%, $(5.2 \pm 0.4)$% for the 7.5%, and $(7.8 \pm 0.4)$% for the 10%. This shift in Mn % from the expected bulk (The bulk Mn composition in our samples was checked by inductively coupled plasma (ICP) analysis [1].) to the surface is shown in Fig. S3. Similar differences between the surface and the bulk were reported [2] for independently grown samples. The distinct size, appearance, number, and location of the impurity sites allows us to differentiate them from other defects intrinsic to the sample.

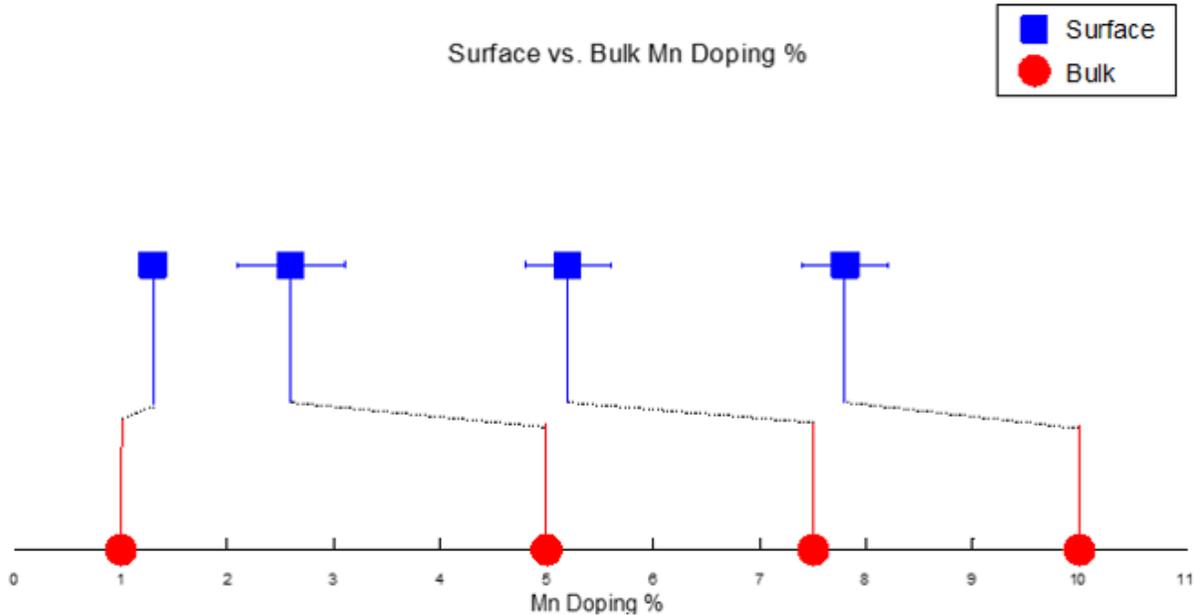

FIG. S3: **Surface vs Bulk Mn doping %.** The expected bulk Mn % (red) and the corresponding observed surface Mn % (blue) are connected by the red and blue lines. There is a clear shift to lower doping on the surface.

## 2. Determination of Gap Width:

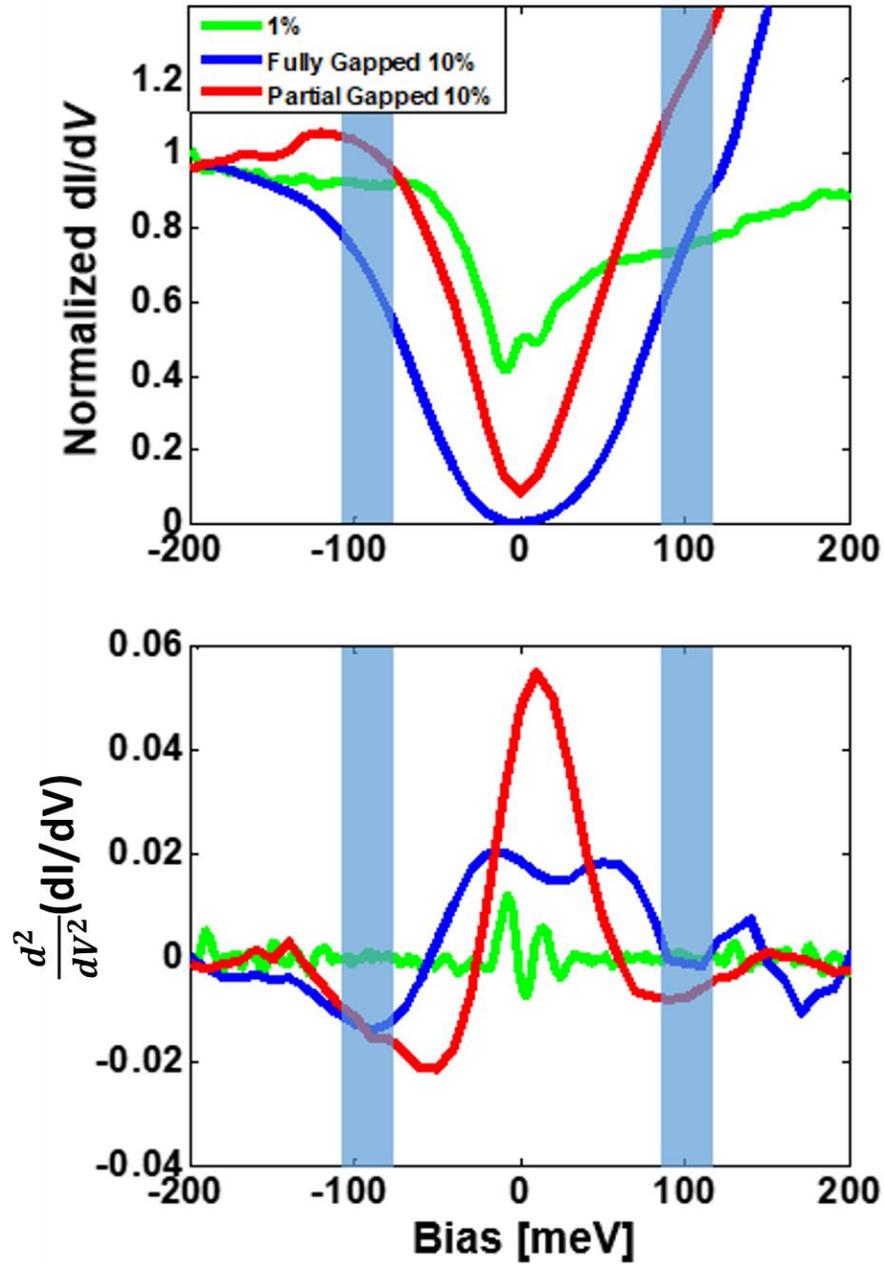

FIG. S4: **Determination of gap width.** Shown in top panel are spectra for a 1%, a fully gapped 10% and a partial gapped 10% taken from the spectra used in fig. 2 of the main text. The bottom panel shows the 2$^{nd}$ derivative of the spectra shown above. The size of the gap can be determined by the local minimum of the second derivative and are marked by the blue area for the fully gapped 10% spectra. All gap sizes used in this work are determined using this method.

## 3. Local effects of Mn impurities on x=1% sample:

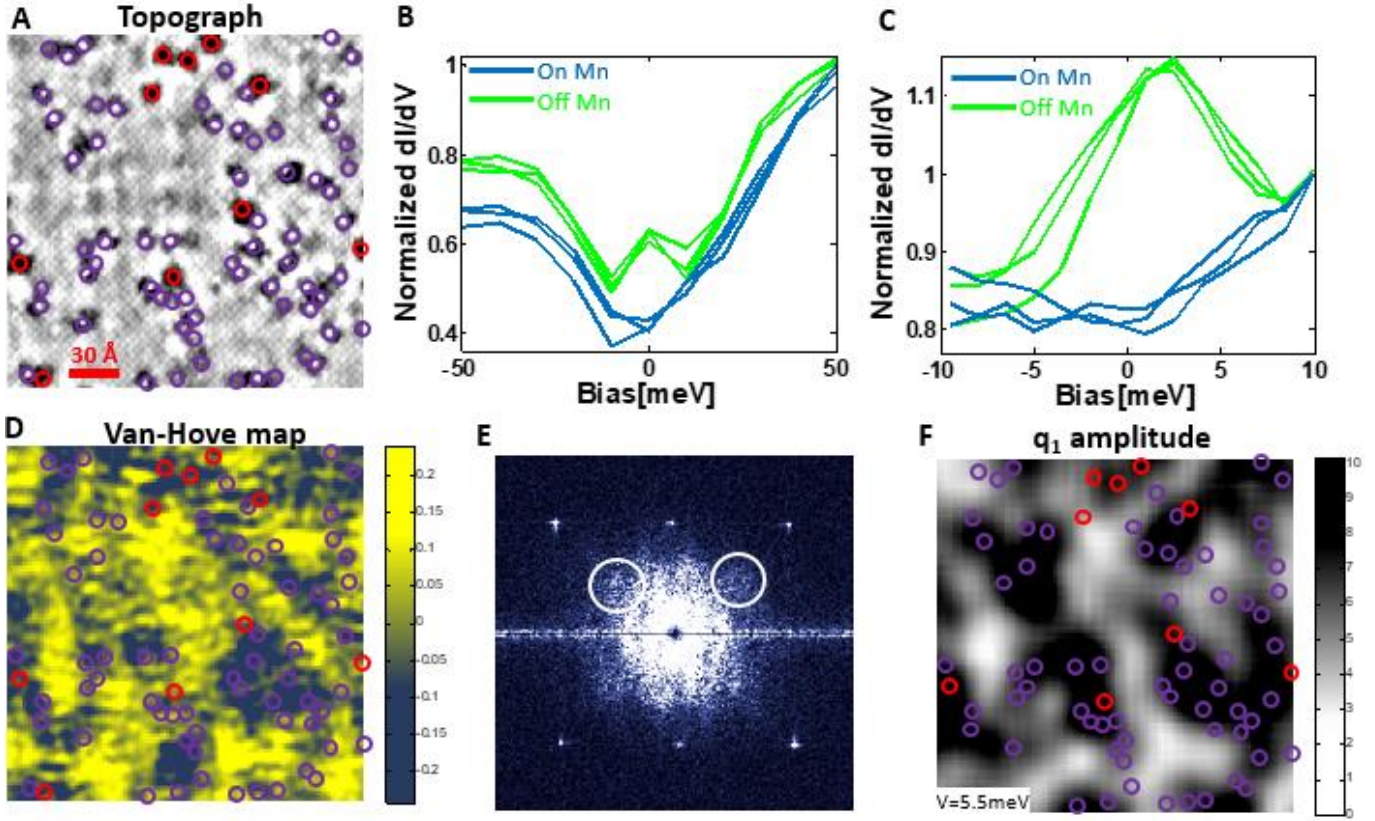

FIG. S5: **Local effects of Mn impurities on x=1% sample** (a) (150mV, 500pA) topography taken on a 1% Mn sample. The purple circles highlight impurity sites. (b),(c) Spectra taken on and off impurity sites, reveal a suppression of the van-Hove singularity near the Mn impurities. (d) Spatial map of van-Hove intensity with impurity overlay show impurity sites to correspond to areas with suppressed van-Hove signal. (e) FT of 1% Mn map with the location of the $q_1$ feature circled. (f) Spatial amplitude of the $q_1$ feature with impurity locations shown.

The local effect of the Mn impurities on the spectra in the 1% doped sample is a local suppression of the density of states near $E_F$ in the vicinity of the van-Hove singularity on impurity sites. Figure S5 (a) shows a topography on a 1% sample with impurity locations marked. The purple circles are the Mn sites described above, and the red circles are defect sites. Spectra on and off impurity sites is shown in Fig S5 (b),(c), revealing a strong feature near $E_F$ on clean areas that is suppressed near impurities. Figure S5 (d), shows a spatial map of the intensity of the hump at $E_F$ (van-Hove map) obtained by plotting the conductance map corresponding to the energy $(2*G(r, 2mV) – G(r,-6.5mv) –G(r, 8.5mV))$. An overlay of impurity sites reveals the impurities to be grouped on areas of suppressed van-Hove peak, with a strong van-Hove signal only appearing on areas free of impurities.

While there is no charge order and no strong Q* peak present in the 1% sample, a weak feature ($q_1$) is still present at the same wavevector. Following a similar procedure described below for the Q* peak, Fig S5 (f), shows a spatial map of the amplitude of this broad feature, with dark areas corresponding to high intensity and light areas to low or zero intensity. Overlaying impurity locations reveals strong intensity locations (corresponding to broad modulations with the q1 wavevector) to correspond to areas of Mn impurities, with zero contribution coming from clean areas, indicating this feature to originate from local effects of the Mn impurities. Cross correlation taken between the van-Hove map and the q1 amplitude map gives a value of -0.43. This large negative value (anticorrelation) indicates that where the van-Hove peak is strong (away from Mn impurities), the q1 amplitude is weak. This analysis indicates that whether the origin of the q1 feature is related to quasiparticle interference off impurities or the beginning of local short-range ordering, it originates from the Mn dopants. In addition, no ordering can be observed locally on areas away from Mn establishing the absence of long range order.

## 4. Raw and Symmetrized FT of STM conductance maps

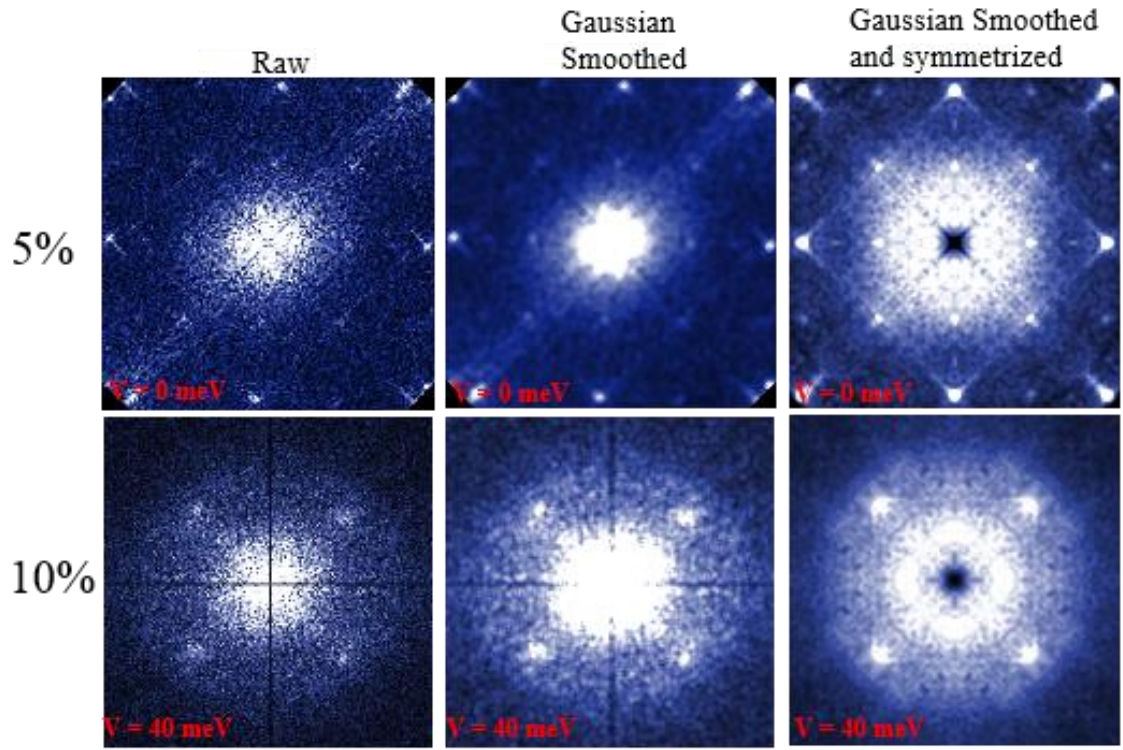

FIG. S6: **Raw and Symmetrized FT of STM conductance maps**. Raw Fourier Transform (FT) of the conductance maps for 5% and 10% Mn samples shown in Fig 3 of the main text (left). To enhance the noise to signal ratio, the FTs are Gaussian smoothed (middle) and four-fold symmetrized (right). The bright spot in the center (very small q) due to long wavelength fluctuations is removed by subtracting Gaussian-filtered conductance map from the original raw conductance value at every pixel. The 2D Fourier transform gets rid of only the low q-vector components. The symmetrized FTs are properly rotated to have the Bragg peaks at the corners of the image.

## 5. Dispersive Measurements of the 1% Mn Sample:

Figure S7 shows line cuts taken along the ($\pi/a$, $\pi/a$) direction for the symmetrized 1% data shown in Fig 3, with their corresponding fits shown as a dashed red line. According to [3], two QPI features are expected to be observed here, namely $q2$ and $q8$. Our linecuts show two sets of dispersive features which we attribute to these two wave vectors. The $q8$ and $q2$ features were fit separately and then combined in the figure. The fits are constructed from a background made of a 2$^{nd}$ order polynomial. The $q8$ feature is fit to a single Gaussian with the q value extracted from the fit as the center of the Gaussian and the error bars as the 95% confidence bounds. The $q2$ feature is fit to two separate Gaussians due to its odd shape (which has no reason to be a simple Gaussian). The q value is extracted as the average of the center of the two Gaussians with the error bars determined by

error=$\sqrt{(error\ of\ first\ Gaussian)^2 + (error\ of\ second\ Gaussian)^2}$ where the error of each

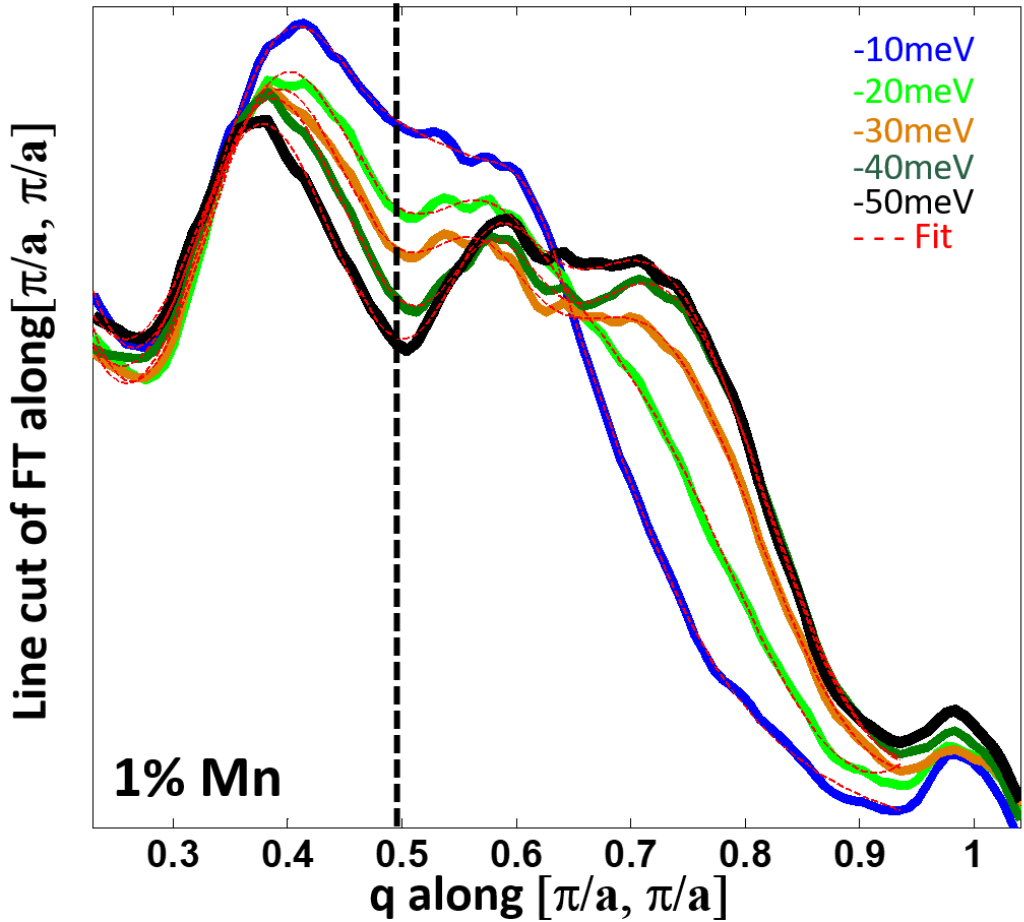

FIG. S7: **Line cuts through the q8 and q2 features** in the 1% Mn data used in Fig 3, with the corresponding fit shown as a red dashed line. Note: The small bump at the end is the satellite reflection peak. The dashed black line at q=0.5 serves to highlight the different between the 1% and 5% samples and the appearance of $q_{AFM}$ with doping in fig.S10 below.

Gaussian is taken as the 95% confidence bounds. There is no reason to expect such a feature to conform to a single equation fit, such as a single Gaussian, indeed when comparing our data to that seen in reference [3], one would not expect a single Gaussian to be sufficient. Despite using two separate Gaussians to fit the q2 feature, we feel comfortable treating it as a single feature, due to the presence of only two total features along the same direction in reference [3].

6. **Non-dispersive behavior of the Charge Order and AFM peaks:**

Figure S8 (b),(e) show line cuts taken from the center through the $Q^*$ peaks in the FTs of a 5% and 10% sample respectively. All data extracted from the 5% and 10% samples was taken from the raw data with the exception of the 5% -100meV Bias $Q^*$ data which was taken from the symmetrized data shown in figure 3. Each line is a different energy from 200mV to -40mV in 20mV intervals for the 5% (The data here were obtained from a conductance map with a setpoint bias of 200meV) and 100mV to -100mV in 10mV intervals for the 10%. The values below -40mV have been excluded as the center inhomogeneity becomes too large and covers up the $Q^*$ peaks. The blue circles are the raw line cut data, while the red lines correspond to a fit using a Gaussian function (for the $Q^*$ peak) and a 3$^{rd}$ order polynomial (as background). (Note: a small peak is present in the 10mV, 0mV, and -10mV cuts for the 10% but its intensity compared to the background is too low for it to be seen clearly. For this reason we have chosen to not include these energies in fig 3(h), as a q value cannot be reliably extracted.) Figure S8 (c),(f) show the raw data and fit for the average of all the energies shown in b,e as well as the raw data and fit for the 40mV in the 5% and the 80mV in the 10%. The green bar is the width of the corresponding Bragg peak for scale. (Note: The FT used for the 5% is taken from a much smaller real space map compared to the 10%. Smaller real space maps yield lower resolution FT and explains why all the peaks in the 5% are much broader even when compared to other 5% peaks such as those seen in the FT in Fig. 3. This map was chosen as the $Q^*$ peak is measured over a larger range of energies.) The q value is extracted from the fit as the center of the Gaussian, the error bars are 95% confidence bounds of the center. Note that there is a percieved shift in the q value for the 10% Mn sample across the Fermi energy, which is due to a discrete change in the background for negative and positive energies as can be seen in Fig S8(e).

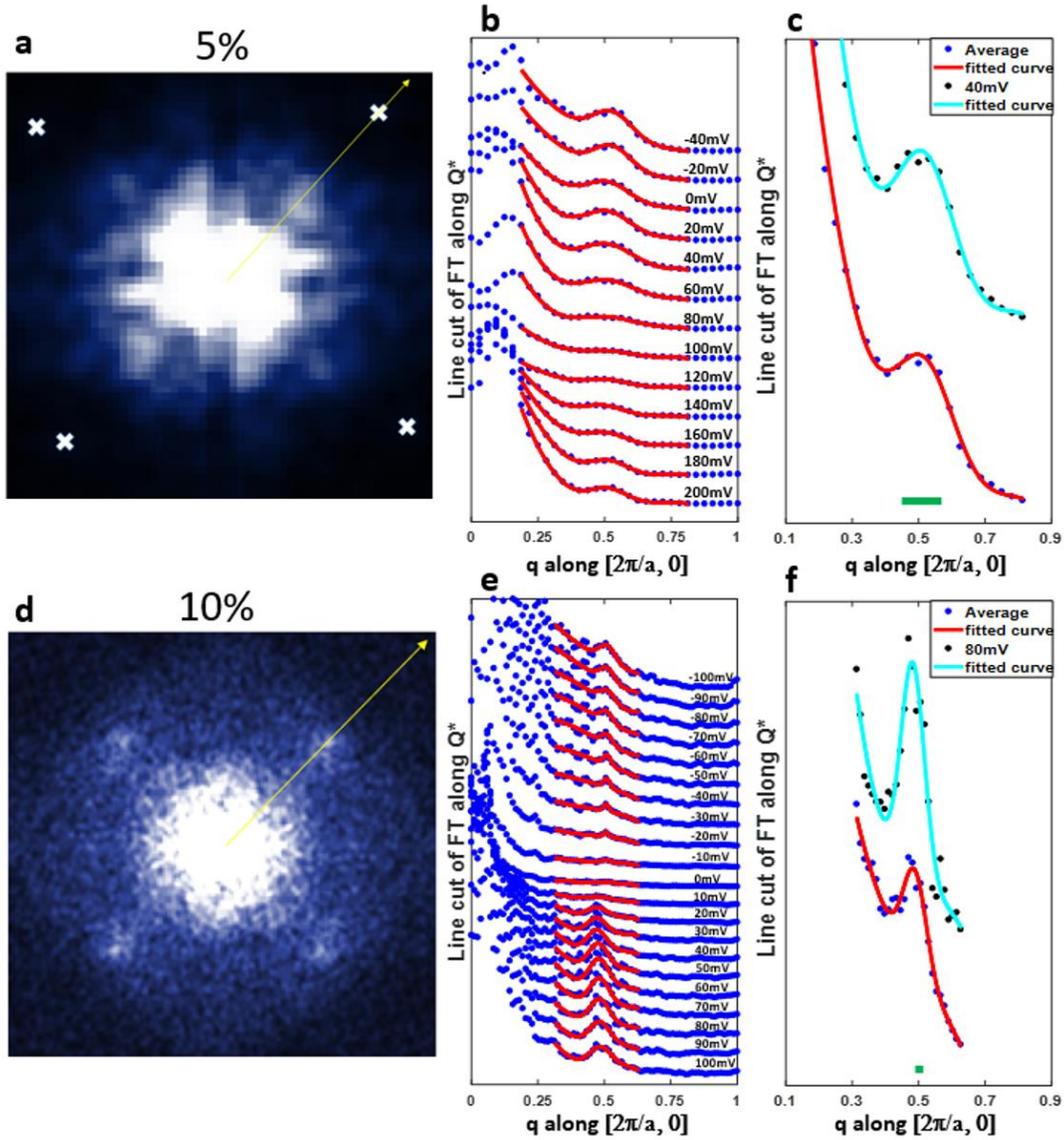

FIG. S8: **Energy dependent measurements of Q\*.** (a),(d) FT of the conductance maps measured on a 5% (size:125Å bias: 60meV) and 10% (size:500 Å, bias:80meV) samples. The yellow arrow marks the path of the line cuts. The white x's denote the position of the Bragg peaks in (a). The Bragg peaks in (d) are placed at the corners. (b),(e) Line cuts taken through the Q\* peak in the 5% and 10% samples respectively. The blue circles are the raw data for each marked energy while the red line shows a 3$^{rd}$ order polynomial plus a Gaussian fit to the data. Lines are shifted upward for clarity. (c),(f) Raw data and fit for the average of all energies and 40mV data for the 5%, and 80mV data for the 10%. The green bar is the full width of the corresponding Bragg peak taken as the FWHM of a Gaussian fit to the Bragg peaks of the corresponding data.

Figure S9 shows additional line cuts taken along the Q* direction showing Q* for the 5% data shown in Fig 3, with their corresponding fits shown as a dashed red line. The data here were obtained from a conductance map with a setpoint bias of -100meV. The fits and extracted values are done using the same method described above for the Q* peak. The values extracted were used for the -100meV values in Fig 3(h).

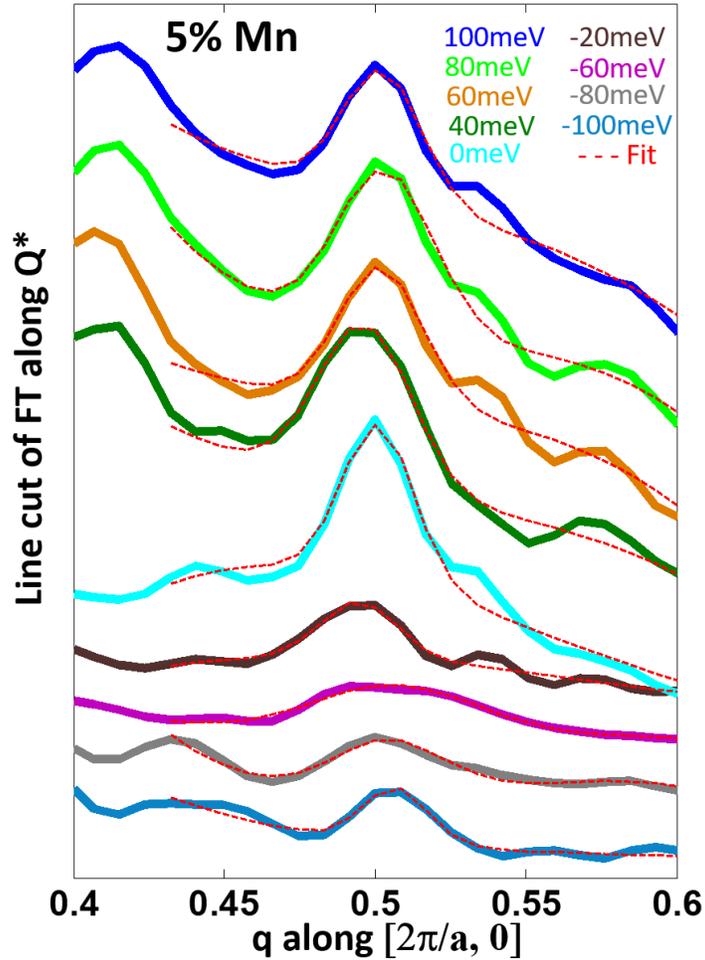

FIG. S9: **Line cuts taken through the Q* peak** in the symmetrized 5% sample shown in Fig 3, with the corresponding fits shown as a dashed red line. The -100meV setpoint Bias is responsible for the suppressed intensity at energies around half of the Bias (-60meV, -80meV, with the -40meV excluded for this reason) due to the setpoint effect. Lines are shifted upward for clarity.

Figure S10 shows similar line cuts taken along the ($\pi/a$, $\pi/a$) direction showing $q_{AFM}$ for the 5% data shown in Fig 3, with their corresponding fits shown as a dashed red line. The fits and extracted values are done using the same method described above for the Q* peak. There may be a dispersive feature present between 0.4 and 0.45 that could be related to q8. This feature which appears to move to higher q with positive energies, along with an enhanced background makes it difficult to extract any values from fits at positive energies.

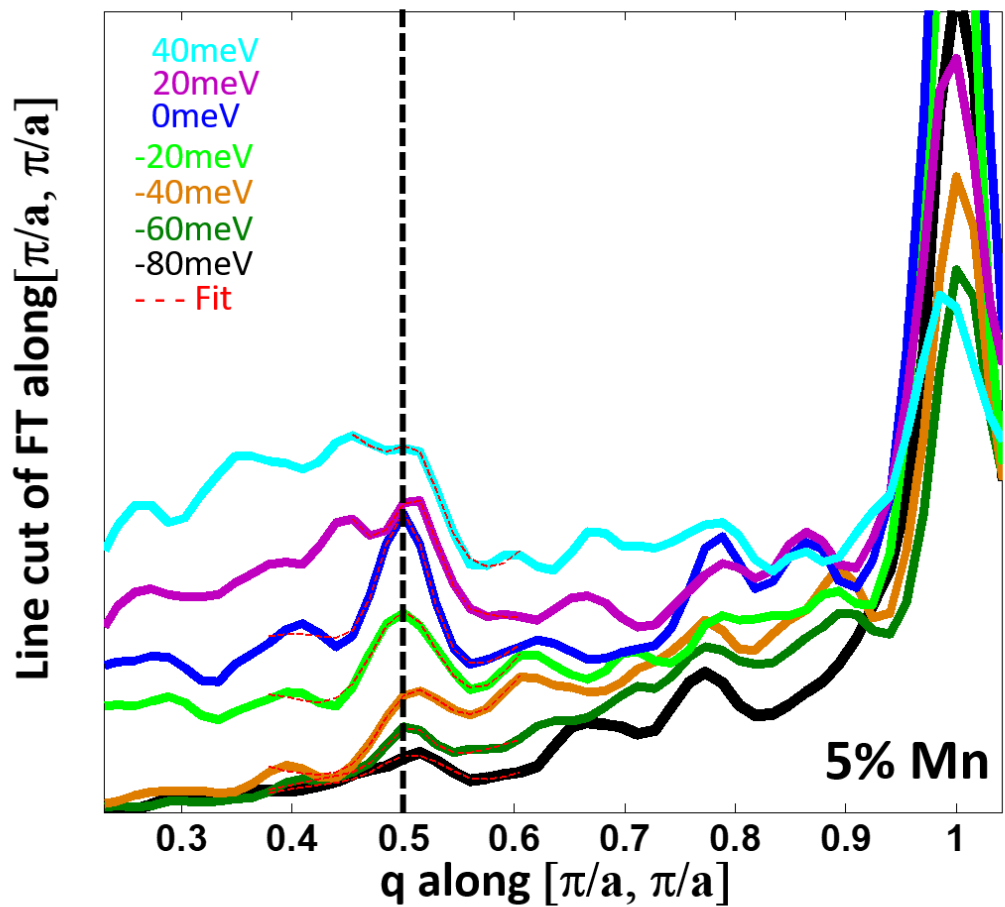

FIG. S10: **Line cuts taken through the $q_{AFM}$ peak** in the 5% sample with the corresponding fits shown as a dashed red line. The dashed black line marks the q=0.5 location. Lines are shifted upward for clarity.

## 7. Theoretical insights into the electronic states of $Sr_3(Ru_{1-x}Mn_x)_2O_7$:

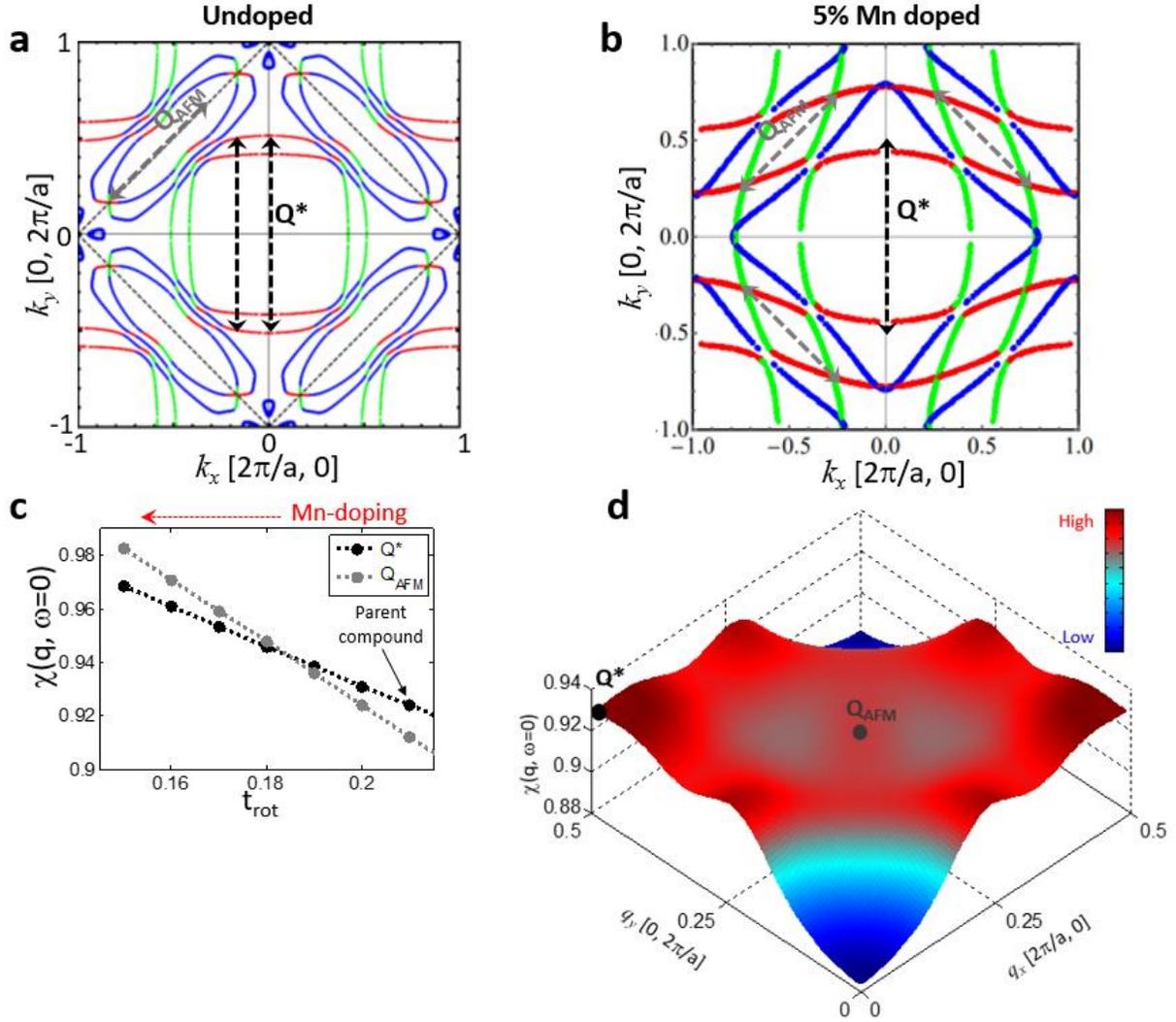

FIG. S11: **Theoretical insights into the electronic states of $Sr_3(Ru_{1-x}Mn_x)_2O_7$** (a) tight binding calculation of the Fermi surface of undoped $Sr_3Ru_2O_7$. (b) Tight binding calculation of the Fermi surface of 5% Mn doped $Sr_3Ru_2O_7$ (c), Evolution of the susceptibility at $Q^*$ and $Q_{AFM}$ with decreasing the $RuO_6$ octahedral rotation. Decreasing the octahedral rotation in the calculations corresponds to increase doping. (d), Real part of the susceptibility for the parent compound showing enhanced peaks at $Q^*$ and a local maximum at $Q_{AFM}$.

## 8. Visualizing the structure and symmetry of Charge ordering patterns:

Figure S12 shows the real space conductance map and its corresponding raw FT of the 10% doped sample used in Fig. 4 of the main text. The Q* peaks corresponding to the charge order (CO) are marked in the FT.

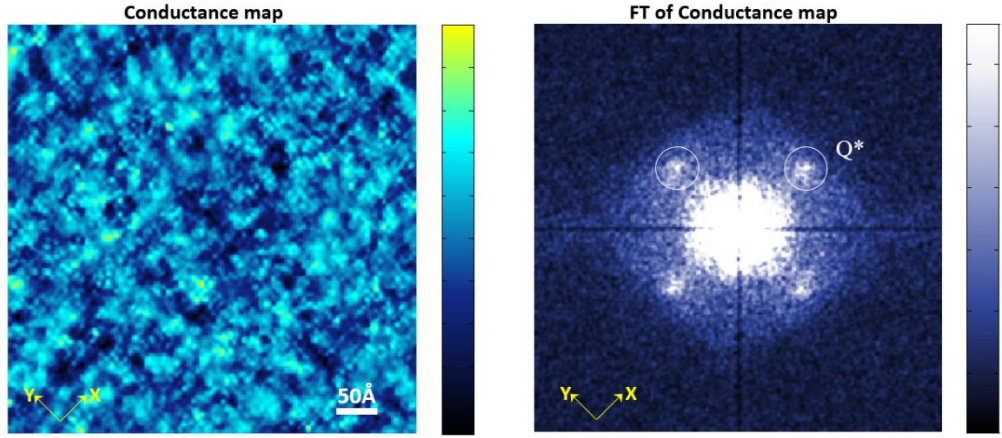

FIG. S12 **Visualizing the structure and symmetry of Charge ordering patterns** Left: Conductance map for the 10% Mn doped sample shown in Fig. 4 of the main text. The long wavelength inhomogeneity in the map is mostly due to the large amount of Mn dopants. Right: Corresponding un-symmetrized FT of the conductance map with the Q* peaks circled.

The FT reveal a broad intensity peak at its center originating from the large amount of inhomogeneity of the real space, mostly due to the large amount of dopants.

To view the CO patterns in real space we multiply the FT in Fig. S12 by broad (with Full width corresponding to the white circles in Fig. S12) Gaussians centered at the four Q* positions. Taking the inverse FT of this reveals the structure of the CO in real space. Figure S13 (a)-(c) shows the results of this method for all the Q* peaks, only the peaks along the Y direction, and only the peaks along the X direction, respectively. These maps reveal a glassy modulation along both directions. The yellow circle in Fig. S13(a) is the expected CO patch size based on the Q* peak in the FT of Fig. S13. The width of the peak is measured as the full width at half maximum (FWHM) of a Gaussian fit to a line cut through the Q* peak this value is than converted into Å ($\lambda=2\pi/\text{FWHM}$) to find the diameter of the circle.

Figure S13 (d)-(f) show the local amplitude of the modulations for the three cases in Fig. S13 (a)-(c). The red areas correspond to large modulating amplitudes, whereas white areas correspond to small or zero amplitude.

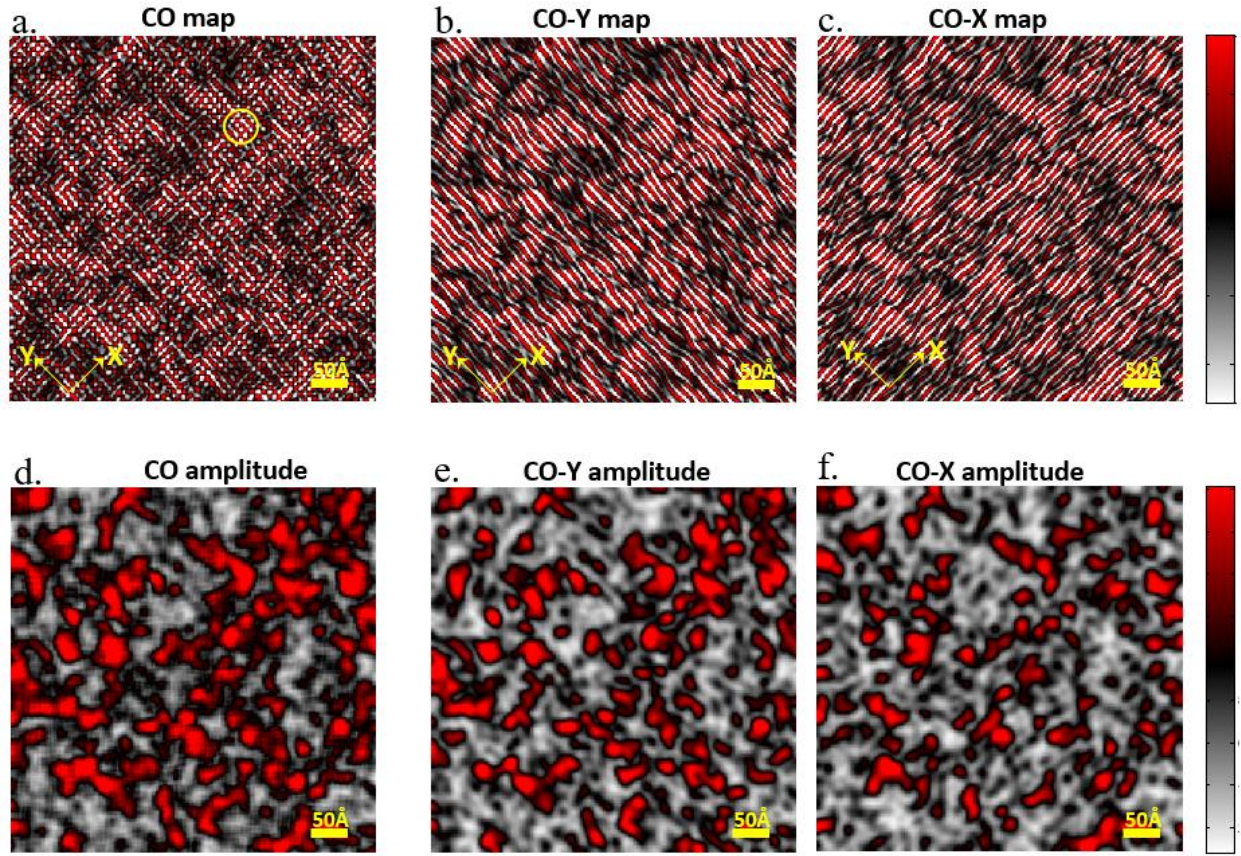

FIG. S13: **Visualizing the structure and symmetry of Charge ordering patterns** (a)-(c) Inverse FTs of the image in Fig. S12 (right) after applying Gaussian filters to isolate all the Q* peaks (a), just the Y direction peaks (b), and just the X direction peaks (c) respectively. (d)-(f) Corresponding amplitude maps of the modulations showing areas of high charge order intensity (red) and areas of low intensity (white). The yellow circle in (a) is the expected patch size based on the width of the Q* peak in the FT in Fig. S12.

Supplementary References:

1. Mathieu, R. et al. Impurity-induced transition to a Mott insulator in $Sr_3Ru_2O_7$. Phys. Rev. B **72**, 92404 (2005).
2. Li, G. R. et al. Atomic-Scale Fingerprint of Mn Dopant at the Surface of $Sr_3(Ru_{1-x}Mn_x)_2O_7$. Sci. Rep. **3**, (2013).
3. Lee, J. et al. Heavy d-electron quasiparticle interference and real-space electronic structure of $Sr_3Ru_2O_7$. Nat Phys **5**, 800 (2009).